\documentclass[twocolumn,showpacs,preprintnumbers,amsmath,amssymb,superscriptaddress]{revtex4}

\usepackage{graphicx}
\usepackage{dcolumn}
\usepackage{bm}
\usepackage{epsfig}

\begin{document}

\preprint{APS/123-QED}

\title{Coulomb suppression of the stellar enhancement factor}

\author{G. G.\,Kiss}%
\affiliation{%
Institute of Nuclear Research (ATOMKI), H-4001 Debrecen, POB.51., Hungary}%
\author{T.\,Rauscher\footnote{Corresponding author, Thomas.Rauscher@unibas.ch}}%
\affiliation{%
Departement Physik, Universit\"at Basel, CH-4056 Basel, Switzerland}%
\author{Gy.\,Gy\"urky}%
\author{A.\,Simon}%
\author{Zs.\,F\"ul\"op}%
\author{E.\,Somorjai}%
\affiliation{%
Institute of Nuclear Research (ATOMKI), H-4001 Debrecen, POB.51., Hungary}%
\date{\today}

\begin{abstract}
It is commonly assumed that reaction measurements for astrophysics should be
preferably performed in the direction of positive $Q$ value to minimize the
impact of the stellar enhancement factor, i.e.\
the difference between the laboratory rate and the actual stellar rate.
We show that the stellar effects can be minimized in the charged particle channel,
even when the reaction $Q$ value is negative.
As a demonstration, the cross section of the astrophysically relevant $^{85}$Rb(p,n)$^{85}$Sr
reaction has been measured by activation
between $2.16\leq E_\mathrm{c.m.}\leq 3.96$ MeV and 
the astrophysical reaction rate for (p,n)
as well as (n,p) is directly inferred from the data. The presented
arguments are also relevant for other $\alpha$- and proton-induced reactions in the $p$ and $rp$
processes. Additionally, our results confirm a previously derived
modification of a global optical proton potential.
\end{abstract}


\maketitle

\textit{Introduction.} Modern nucleosynthesis studies require large reaction networks,
often including hundreds and thousands of nuclei and their respective reactions
with light particles.
Astrophysical reaction rates employed in reaction network calculations are
determined either directly from cross sections or
from the rate for the inverse reaction by applying detailed balance.
The cross sections are either known from experiment or predicted by theory.
Even when a reaction is experimentally accessible,
often astrophysical rates cannot be directly measured.
Excited states are thermally populated in an astrophysical plasma whereas only reactions on the ground state of the
target can be investigated in the laboratory. A measure of the influence of the excited target states is given by the stellar
enhancement factor $f=r_\mathrm{stellar}/r_\mathrm{g.s.}$, defined by
the ratio of the stellar rate to the ground state rate.

\begin{figure}
\resizebox{0.8\columnwidth}{!}{\rotatebox{270}{\includegraphics[clip=]{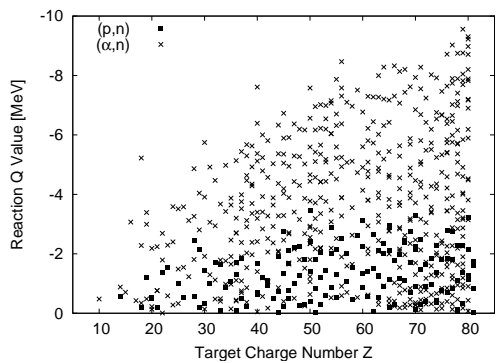}}}
\caption{Reaction $Q$ values for (p,n) and ($\alpha$,n) reactions with $f_\mathrm{rev}<f_\mathrm{forw}$.\label{fig:qdep}
}
\end{figure}

\textit{Relevance of the $Q$ value.}
The enhancement factor $f_\mathrm{rev}$ for the
reverse reaction B(b,a)A (defined by having negative reaction $Q$ value) is usually
larger than the enhancement $f_\mathrm{forw}$ of the forward reaction A(a,b)B
(being the one with positive $Q$ value) because more excited states are energetically accessible in nucleus B than in nucleus A. This is especially pronounced in photodisintegration reactions \cite{UtsZil,moh07}.
Therefore, it was assumed so far that more astrophysically relevant transitions are neglected when experimentally studying a reaction with negative $Q$ value. 
Furthermore, rates employed in reaction networks often use fitted rates
for the forward and backward reaction which are derived from the same rate for one direction to ensure
compatibility and numerical stability. For the latter reason, the fit for the reverse stellar
rate is derived from the one for the forward rate by applying detailed balance. This is the preferred approach because
the relation between backward and forward rate \cite{adndt}
\begin{equation}
r_\mathrm{stellar}^\mathrm{B(b,a)A} \propto e^{-\frac{Q_{\mathrm{A(a,b)B}}}{kT}} r_\mathrm{stellar}^\mathrm{A(a,b)B}
\end{equation}
includes the exponential which would enhance fit inaccuracies when starting from the reverse rate with $Q<0$.

Here, we argue that there are cases for which $f_\mathrm{rev}<f_\mathrm{forw}$
due to Coulomb suppression of a part of the
energetically allowed transitions. This effect will be most pronounced in reactions with a
charged particle in one and a neutral particle in the other channel, e.g.\ (n,p), but it can also
appear when the entrance channel and exit channel have Coulomb barriers of different height, e.g.\ (p,$\alpha$). Transitions from excited states to the same state in a compound nucleus
are proceeding at smaller relative energy and are stronger suppressed by the Coulomb barrier.
Thus, a prerequisite is that $\left| Q\right|$ is low compared to the Coulomb barrier.
Using the NON-SMOKER code \cite{nonsmoker,adndt} we calculated $f_\mathrm{forw}$ and $f_\mathrm{rev}$ for
reactions involving light projectiles (nucleons, $\alpha$) and targets from Ne to Pb between
the proton and neutron driplines. We find more than 1200 reactions exhibiting the suppression
effect. Figure \ref{fig:qdep}
shows the obtained range of $Q$ values as a function of target charge $Z$. It can be clearly seen that
larger $\left| Q\right|$ is allowed with increasing Coulomb barrier. Although the strengths
of the involved transitions also depend on spin and parity of the initial and final state, Coulomb repulsion dominates the suppression when the interaction energy is small.

Not only theoretically interesting, the Coulomb suppression effect is also important for experiments
because it allows to directly determine an astrophysically relevant rate by measuring in the
direction of suppressed enhancement factor. The above mentioned complication of fitting rates
with negative $Q$ values can be circumvented by directly applying detailed balance and numerically
computing the rate for the forward reaction. This is possible when $f_\mathrm{rev}\approx 1$. Subsequently, fits for both rates can be obtained in the standard way.

\textit{Experimental study of $^{85}$Rb(p,n)$^{85}$Sr.}
As an example of the
suppression effect and for the derivation of the astrophysical rates, we experimentally studied the reaction $^{85}$Rb(p,n)$^{85}$Sr, having $Q=-1.847$ MeV. The importance of the reaction is manifold.
In the last several years a number of proton capture cross section measurements with
relevance for $\gamma$ process studies have been carried out 
(see, e.g., \cite{kiss07} and references therein). The $\gamma$ process was shown to synthesize
$p$ nuclides (proton-rich isotopes not accessible to the $s$ and $r$ processes) by a series of photodisintegrations of stable nuclides
in hot layers of massive stars \cite{woo78,arn03,rau02,rau06}.
Recently, systematic
$\gamma$ process simulations found not only that photodisintegration reactions
are important but also that (p,n) reactions, and in particular $^{85}$Rb(p,n)$^{85}$Sr,
strongly influence the final $p$ abundances \cite{rap06}. Additionally, this
reaction is well suited to test
the optical potential used for calculating the
interaction between protons and target nuclei. In low-energy (p,$\gamma$)
reactions, cross sections can be sensitive to the proton as well as the $\gamma$
strength in varying, energy-dependent proportions. Conversely, in (p,n)
reactions the sensitivity is highest to a variation in the proton width
because the neutron width is much larger, thus eliminating the sensitivity
on the width in the exit channel. Compared to the well-studied
(p,$\gamma$) reactions, there is only limited experimental information
available on the low-energy (p,n) cross sections in the mass region of the
light $p$ nuclei. 
The cross section of the $^{85}$Rb(p,n)$^{85}$Sr reaction was already investigated by \cite{kastleiner02} between E$_{c.m.}$ = 3.1 and 70.6 MeV. However, the accuracy is not sufficient for astrophysical applications, mainly because of the large uncertainty of the c.m.\ energies. Moreover, there is only one data point in the relevant energy region for the $\gamma$ process and it bears an uncertainty of $\pm$ 0.5 MeV in the c.m.\ energy. 

\begingroup
\squeezetable
\begin{table}
\caption{$^{85}$Rb(p,n)$^{85}$Sr reaction decay parameters \cite{NDS}}
\setlength{\extrarowheight}{0.1cm}
\begin{ruledtabular}
\begin{tabular}{ccccc}
\multicolumn{1}{c}{Residual} &
\multicolumn{1}{c}{$J^\pi$} &
\multicolumn{1}{c}{Half-} &
\multicolumn{1}{c}{Gamma} &
\multicolumn{1}{c}{Relative $\gamma$-intensity} \\
\multicolumn{1}{c}{nucleus} & &
\multicolumn{1}{c}{life} &
\multicolumn{1}{c}{energy [keV]} &
\multicolumn{1}{c}{per decay [\%]} \\
\hline
$^{85\mathrm{g}}$Sr &9/2$^+$&  64.84 $\pm$ 0.02 d & 514.01 $\pm$ 0.02 & 96 $\pm$ 4 \\
$^{85\mathrm{m}}$Sr$^m$ &1/2$^-$& 67.63 $\pm$ 0.04 m & 231.64 $\pm$ 0.01 & 84.4 $\pm$ 0.2 \\
\end{tabular}
\end{ruledtabular}
\end{table}
\endgroup

\begingroup
\squeezetable
\begin{table*}
\caption{Experimental cross section and $S$ factor values of the
$^{85}$Rb(p,n)$^{85}$Sr reaction.}
\setlength{\extrarowheight}{0.1cm}
\begin{ruledtabular}
\begin{tabular}{ccccc}
\parbox[t]{0.5cm}{\centering{E$_\mathrm{c.m.}$ \\ $\left[ keV \right]$}} &
\parbox[t]{3.8cm}{\centering{Ground state cross section  \\ $\left[mbarn\right]$}} &
\parbox[t]{4.0cm}{\centering{Isomeric state cross section  \\ $\left[mbarn\right]$}} &
\parbox[t]{3.8cm}{\centering{Total cross section  \\ $\left[mbarn\right]$}} &
\parbox[t]{3.8cm}{\centering{ $S$-factor \\ $\left[ 10^6 MeV barn\right]$}} \\
\hline
 2158 $\pm$ 8     & 0.050  $\pm$ 0.005    & 0.008 $\pm$ 0.001    & 0.058  $\pm$ 0.006 & 7.13  $\pm$ 0.67 \\
 2341 $\pm$ 16    & 0.185  $\pm$ 0.013    & 0.039 $\pm$ 0.006    & 0.224  $\pm$ 0.019 & 11.22 $\pm$ 0.96 \\
 2552 $\pm$ 27    & 0.422  $\pm$ 0.029    & 0.147 $\pm$ 0.022    & 0.569  $\pm$ 0.051 & 11.35 $\pm$ 1.02 \\
 2566 $\pm$ 26    & 0.420  $\pm$ 0.032    & 0.162 $\pm$ 0.023    & 0.582  $\pm$ 0.055 & 11.01 $\pm$ 1.04 \\
 2765 $\pm$ 28    & 0.920  $\pm$ 0.087    & 0.282 $\pm$ 0.031    & 1.20  $\pm$ 0.12   & 10.65 $\pm$ 1.05 \\
 2963 $\pm$ 30    & 1.73  $\pm$ 0.17      & 0.391 $\pm$ 0.039    & 2.12  $\pm$ 0.21   & 9.59  $\pm$ 0.92 \\
 3156 $\pm$ 32    & 3.16  $\pm$ 0.29      & 0.612 $\pm$ 0.061    & 3.77  $\pm$ 0.35   & 9.40  $\pm$ 0.87 \\
 3342 $\pm$ 35    & 4.55  $\pm$ 0.40      & 1.11 $\pm$ 0.14      & 5.66  $\pm$ 0.54   & 8.37  $\pm$ 0.79 \\
 3551 $\pm$ 36    & 7.67  $\pm$ 0.64      & 1.93 $\pm$ 0.23      & 9.60  $\pm$ 0.87   & 8.32  $\pm$ 0.76 \\
 3754 $\pm$ 37    & 11.62 $\pm$ 0.90      & 2.69 $\pm$ 0.32      & 14.31 $\pm$ 1.22   & 7.73  $\pm$ 0.66 \\
 3952 $\pm$ 40    & 14.42 $\pm$ 1.18      & 5.23 $\pm$ 0.64      & 19.65 $\pm$ 1.82   & 6.93  $\pm$ 0.64 \\
\end{tabular} 
\end{ruledtabular}
\end{table*}
\endgroup

We measured the $^{85}$Rb(p,n)$^{85}$Sr cross sections using the activation method. Some aspects of the experiment are given here, further details can be found in \cite{kiss07}. Rubidium targets were produced by evaporating natural RbCl
onto a thin Al foil.
The absolute number of target atoms and the uniformity were determined by the Rutherford Backscattering Method (RBS) using the Nuclear Microbeam facility of ATOMKI \cite{simon06}. A 2 MeV He$^+$ beam was focused down to 3x3 $\mu$m$^2$ and was scanned over a surface of 75x75 $\mu$m$^2$ at several different positions of the target. 
Two ion-implanted Si detectors of 18 keV system resolution and 50 mm$^2$ area were applied to collect the backscattering spectra set at $\Theta$ = 165$^{\circ}$ ($\Omega$ = 32.5 msrad) and $\Theta$ = 135$^{\circ}$ ($\Omega$ = 57.3 msrad). The precision of the determination of the number of target atoms was better than 3\%. The thickness was found to be uniform within 1\%.
For the cross section measurements the targets were bombarded with a proton beam of 600 nA typical intensity provided by the Van de Graaff and cyclotron accelerators of ATOMKI. The energy range of the proton beam between 2 and 4 MeV was covered with 200 keV steps. 
To check a possible systematic errors, the E$_p$ = 2.6 MeV irradiation was carried out with both accelerators and no difference in the cross section was found.
Each irradiation lasted approximately 8 hours. 

\begin{figure}
\resizebox{0.75\columnwidth}{!}{\rotatebox{270}{\includegraphics[clip=]{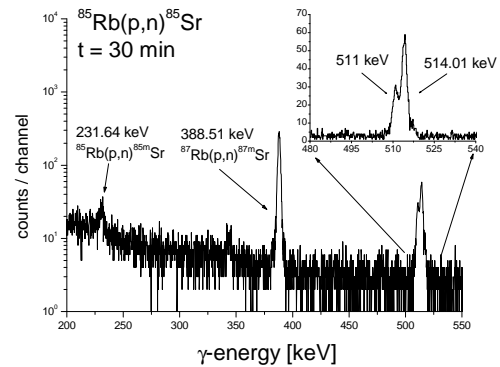}}}
\caption{Off-line $\gamma$ spectrum measured 9 hours after the end of the irradiation
at 2.4 MeV. The 514 keV peak from the $^{85}$Rb(p,n)$^{85}$Sr$^g$ reaction can be well separated from the annihilation peak.\label{fig:spectrum}
}
\end{figure}

The $^{85}$Rb(p,n) reaction populates the ground ($^{85\mathrm{g}}$Sr) and isomeric state ($^{85\mathrm{m}}$Sr) of the Strontium isotope, either
directly or via $\gamma$ cascades.
$^{85\mathrm{g}}$Sr decays by electron capture to $^{85}$Rb and $^{85\mathrm{m}}$Sr with internal transition to $^{85\mathrm{g}}$Sr and with electron capture and $\beta$$^+$ to $^{85}$Rb. For determining the cross section of $^{85}$Rb(p,n)$^{85\mathrm{g}}$Sr the 514 keV, for $^{85}$Rb(p,n)$^{85\mathrm{m}}$Sr the 232 keV $\gamma$ line was used.
The decay parameters of $^{85\mathrm{g,m}}$Sr are summarized in Table I. With the exception of a 388.5\,keV $\gamma$-radiation following the decay of $^{87\mathrm{m}}$Sr, no disturbing $\gamma$-radiations induced on $^{87}$Rb have been observed since all other open reaction channels on $^{87}$Rb lead to stable nuclei.
For measuring the induced $\gamma$ activity a calibrated HPGe detector was used. 
The distance between the surface of the detector and the target was 10 cm. This way the $\gamma$ yield was reasonable and the summing effect was well below 1\%. The yield of the 511 keV annihilation peak (coming from proton induced reactions on target impurities) was less than or comparable to the one of the relevant transition at 514 keV, as shown in Fig.\ \ref{fig:spectrum}. After each irradiation the $\gamma$ spectra were taken for 12 h. Because of the relatively long half life of $^{85\mathrm{g}}$Sr ($T_{1/2}$ = 64.84 d) we were able to repeat the activity measurement for each target after approximately 1 month when the intensity of the 511\,keV radiation is substantially reduced. The two measurements yielded consistent cross sections proving the proper separation of the 511\,keV and 514\,keV peaks. 

The measured total cross sections cover 3 orders of magnitude, varying from 0.058 to 19.645 mb.
Table II lists the measured cross sections $\sigma$ and the $S$ factors, defined as $S(E)=\sigma E^{-1} \exp(-2\pi \eta)$, with the Sommerfeld parameter $\eta$ accounting for the Coulomb barrier penetration \cite{ilibook}.
The quoted uncertainty in the E$_\mathrm{c.m.}$ values corresponds to the energy stability of the proton beam and to the uncertainty of the energy loss in the target. The uncertainty of the cross section values is the quadratic sum of the following partial errors: efficiency of the HPGe detector (6\%), number of target atoms (3\%), current measurement (3\%), uncertainty of decay parameters ($\leq$\,4\,\%) and counting statistics (0.7-4\%).

\begin{figure}
\resizebox{0.7\columnwidth}{!}{\rotatebox{270}{\includegraphics[clip=]{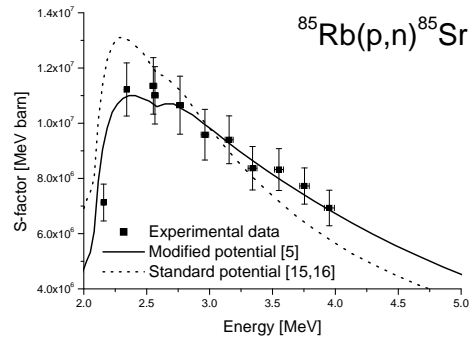}}}
\caption{\label{fig:sfact}Experimental (dots) and theoretical (lines) astrophysical $S$ factors of $^{85}$Rb(p,n)$^{85}$Sr (see text).}
\end{figure}

\textit{Discussion.}
The measured $S$ factors
are compared to theoretical predictions obtained with the code NON-SMOKER 
\cite{nonsmoker,adndt}
in Fig.\ \ref{fig:sfact}. The standard calculation applied a proton optical potential widely
used in astrophysical applications, based on a microscopic approach utilizing a local density approximation \cite{jeu77}.
Low-energy modifications, which are relevant in astrophysics, have been provided by \cite{lej80}. As can be seen in Fig.\ \ref{fig:sfact}, the theoretical energy dependence of the resulting $S$ factor is slightly steeper than the data, although
there is general agreement in magnitude. In the energy range covered by the measurement, the proton width is smaller
than the neutron width (except close to the threshold) and thus uncertainties in the description of the proton width
(and proton transmission coefficient) will fully impact the resulting $S$ factor.
A recent investigation \cite{kiss07} suggested that the strength of the
imaginary part of the microscopic potential should be increased by 70\%.
We find that the energy-dependence of
the theoretical $S$ factor is changed in such a way as to show perfect agreement with the new data,
as seen in Fig.\ \ref{fig:sfact}.
This independently confirms the conclusions of previous work \cite{kiss07}.

Regarding the Coulomb suppression effect, a comparison of $1.03\leq f_\mathrm{pn} \leq 1.08$
and $2.6\leq f_\mathrm{np} \leq 3.9$ shows that the transitions to
excited states of $^{85}$Sr are more important than those to states in $^{85}$Rb
in the relevant plasma temperature range of $2 \leq T \leq 4$ GK.
The almost negligible stellar enhancement $f_\mathrm{pn}$ is due to the suppression of
the proton transmission coefficients
to and from the excited states of $^{85}$Rb
for small relative proton energies because of the Coulomb barrier. There are only few
transitions able to contribute due to the low $Q$ value.
As shown by the small $f_\mathrm{pn}$, the transition from the ground state of $^{85}$Rb dominates the proton channel.
Obviously, a Coulomb suppression is not present in the neutron channel.
On the contrary, for this reaction $f_\mathrm{np}$ is even more enhanced
due to the spin structure of the available nuclear levels and especially the large spin of $^{85}$Sr$^g$.
Because of its large spin, it is connected to the (dominating) low spin states in $^{85}$Rb
through higher partial waves
than the excited states, such as the isomeric state, which have lower spins. Thus, the transitions from the
ground state are suppressed by the centrifugal barrier relative to transitions from excited states and the
latter will quickly become important, even at low temperature. As a consequence of the enhancement of $f_\mathrm{np}$ and the suppression of $f_\mathrm{pn}$, it is more advantageous to measure the (p,n) direction. Important transitions to states
in $^{85}$Sr are included in our data and the small impact of transitions from excited states in $^{85}$Rb
is within the experimental error.

\begingroup
\squeezetable
\begin{table}
\centering
\caption{\label{tab:rates}Astrophysical reaction rates N$_\mathrm{A}$$\left< \sigma v \right>$ of the reactions
$^{85}$Rb(p,n)$^{85}$Sr and $^{85}$Sr(n,p)$^{85}$Rb computed from
experimental data. The values in italics are at temperatures where the experimental data mostly contribute to the rate.
The other values are computed by supplementing theoretical cross sections using the modified optical potential.}
\begin{tabular}{lr@{$\pm$}lr@{$\pm$}l}
\hline \hline
\multicolumn{1}{c}{Temperature}&\multicolumn{2}{c}{$^{85}$Rb(p,n)$^{85}$Sr}&\multicolumn{2}{c}{$^{85}$Sr(n,p)$^{85}$Rb} \\
\multicolumn{1}{c}{[$10^9$ K]}&\multicolumn{2}{c}{[cm$^3$s$^{-1}$mole$^{-1}$]}&\multicolumn{2}{c}{[cm$^3$s$^{-1}$mole$^{-1}$]}\\
\hline
  0.10 & (1.72&0.17)$\times 10^{-89}$ & (1.19&0.2)$\times 10^{4}$ \\  
  0.20 & (8.33&0.83)$\times 10^{-43}$ & (1.74&0.17)$\times 10^{4}$ \\  
  0.40 & (2.26&0.23)$\times 10^{-19}$ & (2.55&0.26)$\times 10^{4}$ \\  
  0.60 & (1.74&0.17)$\times 10^{-11}$ & (3.49&0.35)$\times 10^{4}$ \\  
  0.80 & (1.77&0.18)$\times 10^{-7}$ & (4.80&0.48)$\times 10^{4}$ \\  
  1.00 & (5.07&0.51)$\times 10^{-5}$ & (6.57&0.66)$\times 10^{4}$ \\  
  1.50 & (1.28&0.13)$\times 10^{-1}$ & (1.35&0.14)$\times 10^{5}$ \\ 
{\it  2.00 } & {\it (8.30}&{\it 0.83)} & {\it (2.56}&{\it 0.26)}$\mathit{\times 10^{5}}$ \\  
{\it  2.50 } & {\it (1.21}&{\it 0.12)}$\mathit{\times 10^{2}}$ & {\it (4.57}&{\it 0.46)}$\mathit{\times 10^{5}}$ \\  
{\it  3.00 } & {\it (8.22}&{\it 0.82)}$\mathit{\times 10^{2}}$ & {\it (7.81}&{\it 0.78)}$\mathit{\times 10^{5}}$ \\  
{\it  3.50 } & {\it (3.56}&{\it 0.36)}$\mathit{\times 10^{3}}$ & {\it (1.28}&{\it 0.13)}$\mathit{\times 10^{6}}$ \\  
{\it  4.00 } & {\it (1.15}&{\it 0.12)}$\mathit{\times 10^{4}}$ & {\it (2.04}&{\it 0.20)}$\mathit{\times 10^{6}}$ \\  
  4.50 & (3.03&0.30)$\times 10^{4}$ & (3.17&0.32)$\times 10^{6}$ \\  
  5.00 & (6.89&0.69)$\times 10^{4}$ & (4.76&0.48)$\times 10^{6}$ \\  
  6.00 & (2.60&0.26)$\times 10^{5}$ & (9.52&0.95)$\times 10^{6}$ \\  
  7.00 & (7.14&0.71)$\times 10^{5}$ & (1.54&0.15)$\times 10^{7}$ \\  
  8.00 & (1.50&0.15)$\times 10^{6}$ & (2.01&0.20)$\times 10^{7}$ \\  
  9.00 & (2.50&0.25)$\times 10^{6}$ & (2.18&0.22)$\times 10^{7}$ \\  
 10.00 & (3.44&0.34)$\times 10^{6}$ & (2.05&0.21)$\times 10^{7}$ \\ 
\hline \hline
\end{tabular}
\end{table}
\endgroup

Table \ref{tab:rates} gives the stellar rates for $^{85}$Rb(p,n)$^{85}$Sr as well as for $^{85}$Sr(n,p)$^{85}$Rb. 
Our data cover an energy range sufficient to compute the rates from 2 GK up to 4 GK. Because of the excellent agreement of theory
with experiment, we supplement the data with the theoretical values to compute the rates at lower and higher temperatures, applying
the same errors as for the data. It is to be noted that fits of the rates should be obtained by first fitting the (n,p)
rate and then deriving the (p,n) rate fit by modifying the fit coefficients according to detailed balance (see \cite{adndt}
for details). For convenience, we provide the fit coefficients (including a 10\% error) for the (n,p) rate in the widely used REACLIB format \cite{adndt}: $a_0=16.7791^{+\ln{1.1}}_{+\ln{0.9}}$, $a_1=-7.0325\times 10^{-4}$, $a_2=0.9683$, $a_3=-9.4828$, $a_4=3.1807$, $a_5=-0.3688$, $a_6=2.3328$. The coefficients
for the (p,n) rate are the same, except $a_0^\mathrm{pn}=17.2912^{+\ln{1.1}}_{+\ln{0.9}}$ and $a_1^\mathrm{pn}=-21.4342$. To obtain the (p,n) rate the value obtained with the seven parameter expression has to be multiplied by the ratio of the temperature-dependent partition functions. Details and the required partition functions can be found in \cite{adndt}.

\textit{Summary.} We measured the astrophysically important
reaction $^{85}$Rb(p,n)$^{85}$Sr close above the threshold in the energy range relevant for the $\gamma$ process.
Our measurement confirms a previously derived modification of the global proton optical potential used in
theoretical predictions. Even more importantly, it was shown that in this case it is possible to derive
astrophysical reaction rates for the (n,p) as well as the (p,n)
direction from our (p,n) data despite of the negative reaction $Q$ value due to Coulomb suppression of the stellar
enhancement. A similar argument applies for all
reactions with $Q<0$ but low $\left| Q\right|$ and charged projectiles. Allowing only nucleons,
$\alpha$ particles, and $\gamma$-s as projectiles or ejectiles, this effect
still appears in more than 1200 reactions, including $\alpha$ captures
relevant in the $p$ process \cite{rau06,rap06} and proton captures close to the proton dripline
in the $rp$ process \cite{sch98} and the $\nu p$ process \cite{froh}.

This work was supported by the European Research Council grant 
agreement no. 203175, the Economic Competitiveness 
Operative Programme GVOP-3.2.1.-2004-04-0402/3.0., OTKA (K68801, T49245),
and the Swiss NSF (grant 2000-105328).
Gy.\ Gy.\ acknowledges support from the Bolyai grant.


\begin{thebibliography}{99}
%
\bibitem{UtsZil} H. Utsunomiya, P. Mohr, A. Zilges, and M. Rayet, Nucl.\ Phys.\ {\bf A777}, 459 (2006).
%
\bibitem{moh07} P. Mohr, Zs. F\"ul\"op, and H. Utsunomiya, Eur. Phys. J. \textbf{A 32}, 357 (2007).
%
\bibitem{adndt} T. Rauscher and F.-K. Thielemann, At.\ Data Nucl.\ Data Tables {\bf 75},
1 (2000).
%
\bibitem{nonsmoker} T. Rauscher and F.-K. Thielemann, in {\it Stellar Evolution,
Stellar Explosions, and Galactic Chemical Evolution}, ed.\ A. Mezzacappa (IOP, Bristol
1998), p.\ 519.
%
\bibitem{kiss07} G.\,G.\,Kiss, Gy.\,Gy\"urky, Z. Elekes, Zs. F\"ul\"op, E. Somorjai, T. Rauscher, and M. Wiescher, \prc {\bf 76}, 055807 (2007).
%
\bibitem{woo78} S. E. Woosley and W. M. Howard, \apj Suppl.\ \textbf{36}, 285 (1978).
%
\bibitem{arn03} M. Arnould and S. Goriely, Phys.\ Rep.\ \textbf{384}, 1 (2003).
%
\bibitem{rau02} T. Rauscher, A. Heger, R. D. Hoffman, and S. E. Woosley, \apj \textbf{576}, 323 (2002).
%
\bibitem{rau06} T. Rauscher, \prc \textbf{73}, 015804 (2006).
%
\bibitem{rap06} W. Rapp, J. G\"orres, M. Wiescher, H. Schatz, and F. K\"appeler, \apj \textbf{653}, 474 (2006).
%
\bibitem{kastleiner02} S.Kastleiner, S.M.Qaim, F.M.Nortier, G.Blessing, T.N. van der Walt, and H.H.Coenen, Appl. Radiat. Isot. {\bf 56}, 685 (2002).
%
\bibitem{NDS} H. Siever, Nucl. Data Sheets {\bf 62}, 271 (1991).
%
\bibitem{simon06} A. Simon, T. Cs\'ak\'o, C. Jeynes, and T. S\"or\'enyi, Nucl.\ Instr.\ Meth.\ {\bf B249}, 454 (2006).
%
\bibitem{ilibook} C. Iliadis, {\it Nuclear Physics of Stars} (Wiley, Weinheim 2007).
%
\bibitem{jeu77} J. P. Jeukenne, A. Lejeune, and C. Mahaux, \prc  {\bf 16}, 80
(1977).
%
\bibitem{lej80} A. Lejeune, \prc {\bf21}, 1107, (1980).
%
\bibitem{sch98} H. Schatz \textit{et al}., Phys.\ Rep.\ {\bf294}, 167 (1998).
%
\bibitem{froh} C. Fr\"ohlich, et al, \prl \textbf{96}, 142502 (2006).

 

\end{thebibliography}
\end{document}